\documentclass[preprint,preprintnumbers,amsmath,amssymb]{revtex4-1}

\usepackage{graphicx}
\usepackage{epsfig}
\usepackage{bm}

\def\be{\begin{equation}}
\def\ee{\end{equation}}
\def\bee{\begin{eqnarray}}
\def\eee{\end{eqnarray}}

\begin{document}

\preprint{Preprint: do not distribute}

\title{On the dynamics of vortex modes within magnetic islands.}
\author{W.A.~Hornsby, A.G.~Peeters}
\affiliation{Theoretical Physics V, Dept. of Physics, Universitaet Bayreuth, Bayreuth, Germany, D-95447}
\email{william.hornsby@uni-bayreuth.de}

\author{M.~Siccinio, E.~Poli}
\affiliation{Max-Planck-Institut f\" ur Plasmaphysik, Boltzmannstrasse 2, D-85748
Garching bei M\" unchen, Germany} 

\date{\today}

\begin{abstract}

Recent work investigating the interaction of magnetic islands with micro-turbulence has uncovered the striking 
observation of large scale vortex modes forming within the island structure [W.A. Hornsby {\it et al.}, Phys. Plasmas {\bf 17} 092301 (2010)].  
These electrostatic vortices are found to be the size of the island and are oscillatory.  It is this oscillatory behaviour 
and the presence of turbulence that leads us to believe that the dynamics are related to the Geodesic Acoustic Mode (GAM), and 
it is this link that is investigated in this paper.

Here we derive an equation for the GAM in the MHD limit,  in the presence of a magnetic island modified three-dimensional 
axisymmetric geometry.   The eigenvalues and eigenfunctions are calculated numerically and then utilised to 
analyse the dynamics of oscillatory large-scale electrostatic potential structures seen in both linear and non-linear gyro-kinetic 
simulations.

\end{abstract}

\maketitle

Magnetic islands, generated by the tearing mode, have the effect of breaking the axisymmetric properties 
of the equilibrium, and can have a detrimental effect on the plasma confinement due to the radial 
component of the magnetic field that they introduce\cite{FUR63,RUTH73,WAE09}.  

Recent work on the interaction of turbulence with large magnetic islands has uncovered the 
striking observation of large scale electrostatic vortices forming within the island 
separatrix (See Fig.~\ref{islpotturb}).  These meso-scale potential structures generate $E\times B$ 
flows around the island, similar to the zonal flows which act as regulators of turbulence\cite{DiaZF,HasZF,WalGF,LinZF,WangVort,WangFlow,Mura09}, having 
the effect of tearing up radially extended electrostatic eddies, and thus reducing the radial transport of particles and heat.     
On the contrary, it is seen that these vortices can enhance the heat flow within the separatrix by up to 50\% by acting as a convective cell, having a detrimental impact on heat confinement in 
a toroidal plasma\cite{HorEPL,POL09,POL10}, but also have a significant effect on the radial pressure profile, which in turn can 
effect the bootstrap current profile which determines the stability of the Neoclassical Tearing mode (NTM)\cite{Hor10,HorVar,CAR86,WIL96} .

Zonal flows in fusion devices are intrinsically linked to an oscillatory mode known as the Geodesic Acoustic 
Mode (GAM)\cite{Win68,Hag09,Shi05}.  
These can be excited non-linearly by primary instabilities such as drift waves  or by interaction with energetic 
particles in fusion plasmas\cite{QiuZon}.
GAMs are generated on closed flux surfaces when perturbations in the $E\times B$ flow couple 
to axisymmetric pressure perturbations, by way of the curvature in the magnetic field causing a 
compression, to produce oscillatory electrostatic modes.    Geodesic Acoustic modes have been 
extensively observed in a variety of laboratory plasmas\cite{ConwayGAM05,HamadaGAM05,MelnikovGAM06,ZhaoGAM06,FujisawaGAM07}.  
The GAM was initially identified in toroidal symmetric systems, however, recently both theory and experiment has extended to 
include helical systems\cite{WatariGAM05,WatariGAM06}.  
It is thought that plasma compressibility, which is the cause of the GAM, can have a significant 
effect on the growth rate of tearing instabilities\cite{SmolGAM}.   

\begin{figure*}
\centering
\includegraphics[width=10.5cm,clip]{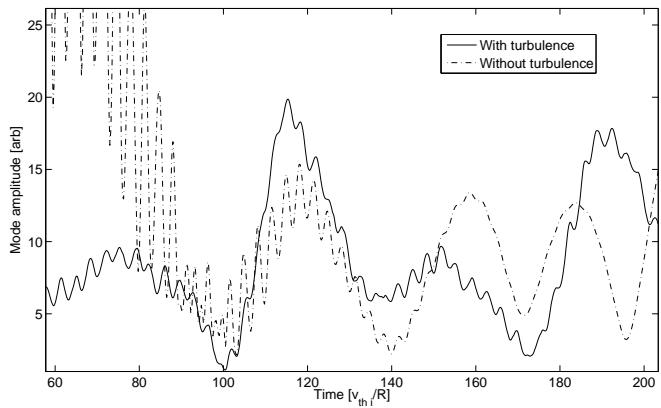}
\caption{Time trace of the amplitude of the smallest radial wave-vector (sideband) as a function of 
time for an island with poloidal wave-vector, $k_\theta \rho_i = 0.05$.  There are two frequencies present, 
the fast, quickly damped Geodesic Acoustic Mode and the slower, island Geodesic mode, which is very slowly 
damped and of larger amplitude.  The amplitudes of the modes have been adjusted so that a direct comparison 
can be made.  Without the presence of turbulence the mode is initialised.  When turbulence is present, the vortex mode is generated.}
\label{tracecomp}
\end{figure*}

\begin{figure}
\centering
\includegraphics[width=8.5cm,clip]{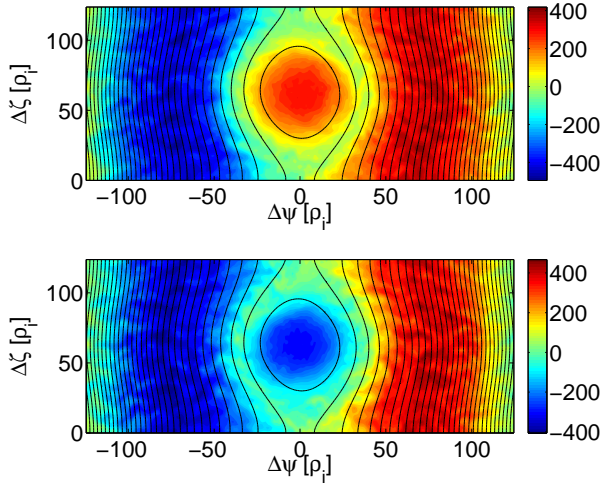}
\caption{(Color online) Normalized electrostatic potential ($\phi_{N}=e\phi/T\rho_*$, $w/\rho_{i}=24$) in the plane perpendicular to the magnetic field (outboard mid-plane).  Black lines represent the perturbed flux surfaces calculated from the total parallel vector potential.  The presence of the island embedded in the turbulence not only generates flows around the island structure but also large scale electrostatic potential structures within the island which fluctuate in amplitude and sign.  The top panel shows a vortex with a positive sign, while the lower panel shows that the vortex has flipped sign at a point later in the simulation.}
\label{islpotturb}
\end{figure}

It is the oscillatory electrostatic structure of the GAM, and the regular oscillation period of the vortex seen in nonlinear simulations (See 
black trace in Fig.~\ref{tracecomp} and 2D slices in Fig.~\ref{islpotturb}), that leads us to believe that similar physics is 
responsible for the oscillatory vortex structures seen inside magnetic islands and it is this observation 
that is the basis of this paper. 

The period of the GAM frequency is determined by the compressibility of the plasma and is closely 
related to the sound speed and to the shape of the flux surface.  The radial component of the magnetic field introduced 
by the tearing mode, produces magnetic islands which are seperate confinement regions within the plasma.
Within these new confinement regions the GAM period is likely to be highly modified due to their helical structure.

The paper is structured as follows.  In sections \ref{model} and \ref{eigenchap}, the model 
is outlined and the Eigenfunctions and frequency of the Geodesic Acoustic mode in the presence of a magnetic island structure are calculated and analysed in section \ref{eigenchap2}.  

Sections \ref{theory} and \ref{compare} will outline 
Gyro-kinetic simulations to study these dynamic structures and then a comparison is made with oscillations 
seen in fully nonlinear gyro-kinetic turbulence simulations with magnetic islands.

\section{Mathematical model}
\label{model}

We begin by calculating the eigenvalue equation for the geodesic acoustic mode in the ideal MHD limit,  and modify the theory to take into account the change in the flux surfaces due to the presence of a magnetic island.  We assume here that the evolution of the island is significantly slower than oscillation time of the mode.  The islands that we
consider are large (an island half-width of $w/\rho_i = 24$),  and are treated as a static structure, rather than a dynamic mode and behave as a separate confinement region within the plasma\cite{biancalani}.

For the equilibrium magnetic field we utilise the axisymmetric, large aspect ratio toroidal geometry, with a further assumption that the flux surfaces are circular.  Here, $\epsilon = r/R$ is the inverse aspect ratio, where r is a minor and R is the major radius.

The helical angle, associated with the magnetic island mode is defined as:
\begin{equation}
\xi = m\theta - n\zeta - \omega t
\end{equation}
where $\theta$ and $\zeta$ are the poloidal and toroidal angles respectively, and m and n are the poloidal and toroidal mode-numbers.  $\omega$ is the island rotation frequency, which in this analysis is set to zero.

The perturbation due to the magnetic island consists of a helical flux component,
\begin{equation}
\psi = \tilde{\psi}\cos{\xi}
\end{equation}
$\tilde{\psi}$ is treated as a constant in accordance with the well utilised constant$-\psi$ approximation\cite{FUR63}. 

With the island present, it is possible to construct modified flux-surfaces with $\Omega$ as the island flux-surface label.  This has the form\cite{FIT95}:
\begin{equation}
\Omega = -\frac{\psi}{\tilde{\psi}} = 2\frac{(r-r_{0})^{2}}{w^2} - \cos{\xi}
\end{equation}
Where $r_{0}$ is the minor radius at the rational surface on which the island is sited and $w$ is the island half-width.  

We are interested, primarily, in modes with a small poloidal wave-vector in the electrostatic limit, so we keep our analysis to the ideal MHD equations. 

Here we present the linearised MHD equations and closely follow the procedure to calculate the Geodesic acoustic mode dispersion relation that is originally outlined by Windsor \textit{et. al}\cite{Win68}.  The equations are as follows:
\begin{eqnarray}
\label{MHDeq}
\rho\frac{\partial \mathbf{\tilde{v}}}{\partial t} &=& \mathbf{\tilde{J}}\times\textbf{B} - \nabla\tilde{p}\nonumber\\
\frac{\partial \tilde{\rho}}{\partial t} &+& \rho\nabla.\mathbf{\tilde{v}} =  0 \nonumber\\
\nabla\tilde{\phi} &=& \mathbf{\tilde{v}}\times\mathbf{B}\nonumber\\
\nabla\cdot\mathbf{\tilde{J}} &=& 0 \nonumber\\
\rho^{-\gamma}\frac{\partial \tilde{p}}{\partial t} &-& \gamma p \rho^{-\gamma-1}\frac{\partial \tilde{\rho}}{\partial t} + \tilde{\mathbf{v}}.\nabla(p\rho^{-\gamma}) = 0
\end{eqnarray} 

Where the equations are the linearised momentum, continuity, Ohms law, quasi-neutrality condition and the equation of state, respectively.  The tilde denotes a perturbed quantity, $\rho$, $\mathbf{\mathbf{J}}$ and $p$ denoting the mass density, current density and pressure respectively.  $\mathbf{B}$, $\phi$ and $\mathbf{v}$ are the magnetic fields, electrostatic potential and the plasma velocity.  $\gamma = 5/3$ denotes the adiabatic constant.  

The flux surface label, $\Omega$, satisfies the magnetic differential equation,
\begin{equation}
\textbf{B}\cdot\nabla\Omega = 0,
\end{equation}
when a magnetic island is present.  With this in mind, the fluid velocity has three components and can be written as:
\begin{equation}
\mathbf{\tilde{v}}= \exp{(-\imath\omega t)}\left( \tilde{v}_{\Omega}\frac{\nabla\Omega}{|\nabla\Omega |^{2}} + \tilde{v}_{\zeta}\frac{\mathbf{B}\times\nabla\Omega}{B^2} + \tilde{v}_{s}\frac{\mathbf{B}}{B^2} \right)
\end{equation}
The first term being the velocity across the flux surfaces, second, the velocity in the binormal coordinate and the last term being the velocity directed along the magnetic field.  From  Ohm's law in Eq.(~\ref{MHDeq}), it follows that the electrostatic potential is just a function of the flux surface label, $\Omega$.
\begin{equation}
\mathbf{\tilde{v}}_{\zeta} = \frac{\mathbf{B}\times \nabla \Omega}{B^{2}}\frac{\partial\phi}{\partial\Omega}
\end{equation}
It can be shown that the term across the flux-surfaces,  $ \tilde{v}_{\Omega}$ is zero and $v_{\zeta} = \frac{\partial\phi}{\partial\Omega}$ which is a flux surface quantity.  This reduces the equation of state to $\tilde{p} = \frac{\gamma p}{\rho}\tilde{\rho}$, the equilibrium pressure given by, $p = \rho k_{B}T/m_{i}$.

Taking the flux surface average of the linearised momentum equation yields in the radial direction,
\begin{equation}
\tilde{v}_{\zeta} = -\imath \frac{\gamma p}{\omega\rho^2} \int \frac{\mathbf{B}\times\nabla\Omega\cdot\nabla\tilde{\rho}}{B^2}JdS / \int \frac{|\nabla\Omega|^{2}}{B^2} JdS,
\end{equation}
where $J$ is the Jacobian, whereas the parallel component can be used to obtain,
\begin{equation}
\tilde{v}_{s} =  -\imath \frac{\gamma p}{\omega\rho^2} \mathbf{B}\cdot\nabla\tilde{\rho}.
\end{equation}
Substitution into the continuity equation we obtain the following Eigenvalue equation for $\omega^{2}$, the squared mode frequency,

\begin{eqnarray}
\omega^{2}\tilde{\rho} = -\left(\frac{\gamma p}{\rho}\right)\frac{\frac{\mathbf{B}\times\nabla \Omega\cdot\nabla B^{2}}{B^{4}}}{\int \frac{\left|\nabla\Omega\right|^2}{B^2} JdS}& &\int \tilde{\rho} \frac{\mathbf{B}\times\nabla \Omega\cdot\nabla B^{2}}{B^{4}}JdS \nonumber\\
&-&B\nabla_{\parallel}\left(\frac{\nabla_{\parallel}\tilde{\rho}}{B}\right).
\label{eigeneq}
\end{eqnarray}

The first term representing the effect of E-cross-B flows associated with the compression caused by the geodesic 
curvature within the magnetic island, while the second term represents the motion of sound waves parallel to the magnetic field lines.

Multiplying the continuity Eq.(~\ref{MHDeq}) with the complex conjugate of the mass density ($\rho^{*}$) and taking the flux surface average, we arrive at an integral equation for the dispersion relation analogous to the one given in\cite{Win68}:

\begin{eqnarray}
\omega^{2}\int |\tilde{\rho}|^2 JdS = \nonumber\\
\left(\frac{\gamma p}{\rho}\right) \frac{1}{\int \frac{\left|\nabla\Omega\right|^2}{B^2} JdS}\left(\left| \int \tilde{\rho} \frac{\mathbf{B}\times\nabla \Omega\cdot\nabla B^{2}}{B^{4}} JdS \right|^2 \right)\nonumber\\
+\int\left|\nabla_{\parallel}\tilde{\rho}\right|^{2} JdS
\label{disp}
\end{eqnarray}

In the above derivation, finite gyro-radius effects have been neglected and as such does not provide any information about the structure 
of the vortex mode perpendicular to the perturbed flux-surfaces $\Omega$.  This would require a kinetic or higher order approach\cite{sug06}
and is beyond the scope of this paper.

\section{Eigenvalue calculation}
\label{eigenchap}

We calculate the Eigenvalues and Eigenfunctions numerically by writing Eq.(~\ref{eigeneq}) in the form:

\begin{eqnarray}
\omega^{2}\tilde{\rho} &=& -\left(\frac{\gamma p}{\rho}\right)\frac{1}{\int \frac{\left|\nabla\Omega\right|^2}{B^2} JdS} \frac{\mathbf{B}\times\nabla \Omega\cdot\nabla B^{2}}{B^{4}} C  \nonumber\\
&-&B\nabla_{\parallel}\left(\frac{1}{B} \nabla_{\parallel}\tilde{\rho}\right)
\label{eigen}
\end{eqnarray}
\noindent
where:
\begin{equation}
C = \int \tilde{\rho} \frac{\mathbf{B}\times\nabla \Omega\cdot\nabla B^{2}}{B^{4}}JdS
\end{equation}

which can be written in the form of a generalised Eigenvalue equation and then solved using 
standard methods.

In magnetic island geometry\cite{WIL09,sicc09,jame,smol95}, $\nabla_{||}$ is defined as:
\begin{equation}
\nabla_{\parallel} = \frac{1}{Rq}\frac{\partial}{\partial\theta}\Bigg|_{\Omega} + k_{\parallel}\frac{\partial}{\partial\xi}\Bigg|_{\Omega}
\end{equation}
\noindent
and remembering that $k_{\parallel} = -k_{\theta}(r-r_0)/L_s$ and also $k_{\theta} = nq/r = m/r$ and the shear length, $L_{s}$ is defined as, $L_{s} = Rq/\hat{s}$\cite{sicc09,jame,smol95} 
\begin{equation}
k_{||} = \mp\frac{w m \hat{s}\sqrt{(\Omega + \cos{\xi})}}{\sqrt{2}q \epsilon R}
\end{equation}
\noindent
Where $r_0$ is the radial coordinate of the rational surface of consideration, the negative sign is chosen when  
$(r-r_0)$ is positive and vice-versa.

We note here that the flux surface integral is defined by firstly taking the integral over the poloidal angle $\theta$
then by an integral over the helical angle $\xi$. 
For an arbitrary function, A, this flux surface average can be written as\cite{WIL96,FIT95},
\begin{equation}
\langle A \rangle  =  \int  \frac{Ar_{0}(1+\epsilon\cos{\theta})}{\sqrt{\Omega + cos{\xi}}} d\theta d\xi.
\end{equation}

Consider the term $\int \frac{\left|\nabla\Omega\right|^2}{B^2}JdS$.
Firstly we note that, we use the simple circular cross-section, axisymmetric model for tokamak equilibrium, with the approximations, $\nabla r = \hat{\mathbf{r}}$,$\nabla\theta \sim\frac{\hat{\mathbf{\theta}}}{r}$ and
$\nabla\zeta \sim \frac{\hat{\mathbf{\zeta}}}{R}$.

If we utilise:
\begin{equation}
\nabla\Omega = \frac{4(r-r_0)}{w^2}\hat{\mathbf{r}} + \frac{m}{r}\sin{\xi}\hat{\mathbf{\theta}} - \frac{n}{R}\sin{\xi}\hat{\mathbf{\zeta}}
\end{equation}
Performing the $\theta$ integral, which removes the terms of order $\epsilon$.
We assume a small aspect ratio $(r<R)$ and also small island width in relation to the minor radius$(w<r)$ and as such the integral reduces to:
\begin{equation}
\int \frac{\left|\nabla\Omega\right|^2}{B^2}JdS = \frac{32\pi}{B_0^2 w^2} \int \frac{(\Omega + \cos\xi)}{\sqrt{\Omega + \cos\xi}} d\xi
\label{part3}
\end{equation}
\noindent
Finally we treat the term $\int \tilde{\rho} \frac{\mathbf{B}\times\nabla \Omega.\nabla B^{2}}{B^{4}} JdS$, for brevity the full details of this calculation can be found in Appendix A.  
Utilising $\mathbf{B} = B_{t}\hat{\mathbf{\zeta}} + B_{\theta}\hat{\mathbf{\theta}} + \frac{\tilde{\psi}m}{Rr}\sin\theta \hat{\mathbf{r}}$, neglecting the effect of the island on the field strength, utilising:
\begin{eqnarray}
B_t &=& B_0/(1+\epsilon\cos\theta)\\
B_\theta &=& \frac{\epsilon B_0}{q}/(1 + \epsilon\cos\theta) 
\end{eqnarray}
\noindent
and the major radius varying according to:
\begin{equation}
R = R_{0}(1 + \epsilon\cos\theta),
\end{equation}

the compression term $\frac{\mathbf{B}\times\nabla B^{2}\cdot\nabla\Omega}{B^{4}}$ can be written as:

\begin{eqnarray}
\frac{\mathbf{B}\times\nabla B^{2}\cdot\nabla\Omega}{B^{4}}&=& \frac{-2}{B_{0}R}\Bigl(\frac{4\sqrt{\Omega+\cos\xi}}{w\sqrt{2}}\sin\theta \nonumber\\ 
&+&\Bigl(\frac{m}{r}+\frac{\epsilon n}{R_{0}q}\Bigr)\cos\theta\sin\xi  \Bigr)
\end{eqnarray}

This equation is substituted into 
Eq.~(\ref{eigen}), which forms a generalised eigenvalue equation.  Both positive and negative $r-r_0$ sides of the magnetic island are considered with periodic boundary 
conditions and the eigenfunctions and eigenvalues calculated.  The results are discussed in next section.

Far away from a magnetic island, where the perturbation of the magnetic flux surfaces is smallest, 
it is expected that the dispersion relation will return to the form of the standard Geodesic acoustic mode.
 
Taking the limit of large $\Omega$ and large aspect ratio, $\epsilon < 1 $.  Taylor expanding then performing the integrals
 we obtain:

\begin{equation}
\omega^{2} = \frac{\gamma v_{th}^{2}}{2R^{2}}\left(2 + \frac{1}{q^{2}} \right),
\end{equation}

From Fig.~\ref{tracecomp}, we see that the high frequency GAM is present in both the simulations with and without electrostatic turbulence, having an identical frequency in both.

\section{Eigenfunction Analysis}
\label{eigenchap2}

\begin{figure}
\centering
\includegraphics[width=8.5cm,clip]{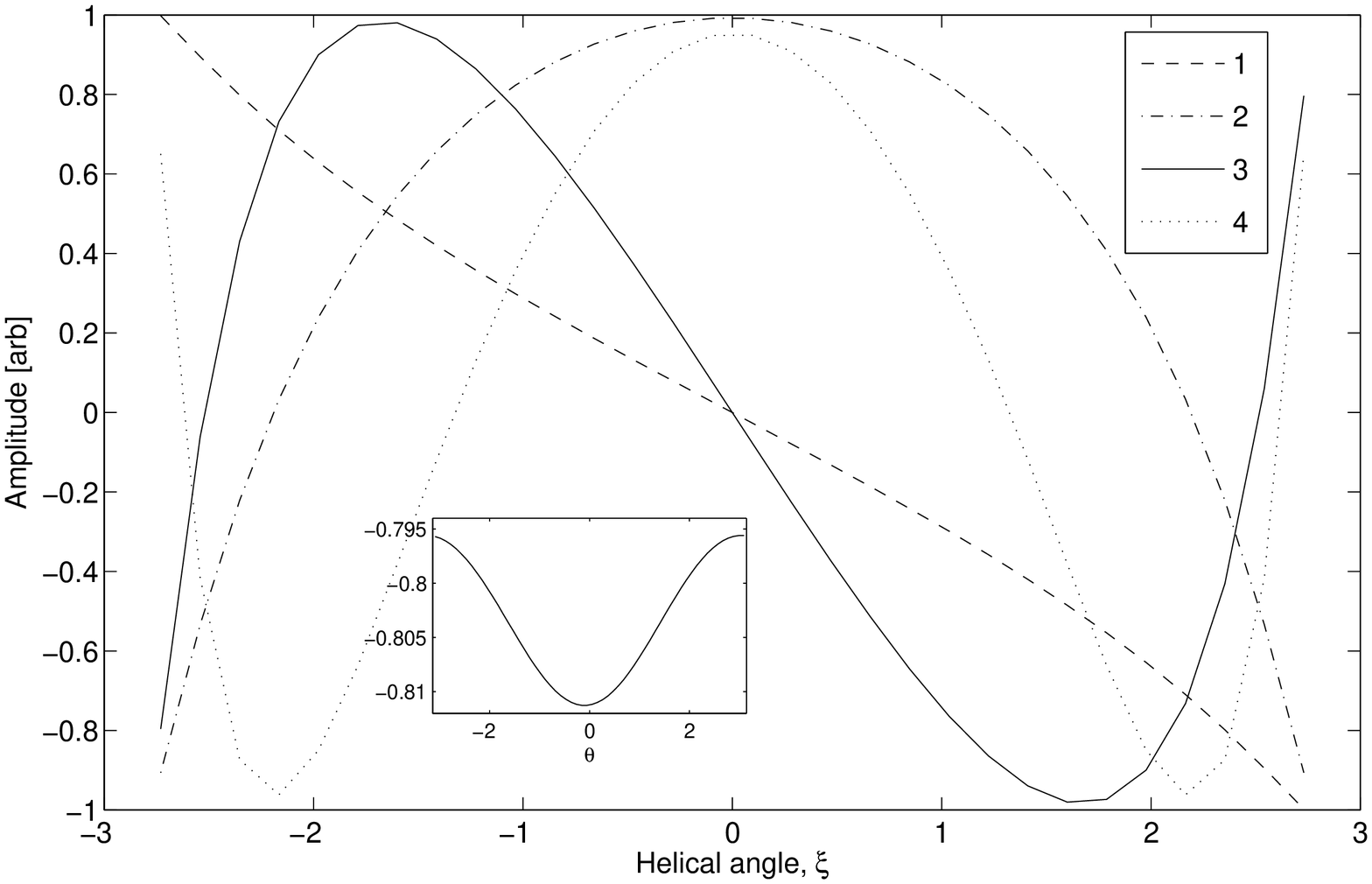}
\caption{The non-trivial density Eigenfunctions in the direction of the helical angle, $\xi$ at $\Omega=0.95$. Shown in inlay is the Eigenfunction in the poloidal angle.  Plotted here is one half of the island eigenfunction (i.e. positive or negative radial direction).  The curves are numbered according to increasing Eigenvalue.}
\label{eigenfunction}
\includegraphics[width=8.5cm,clip]{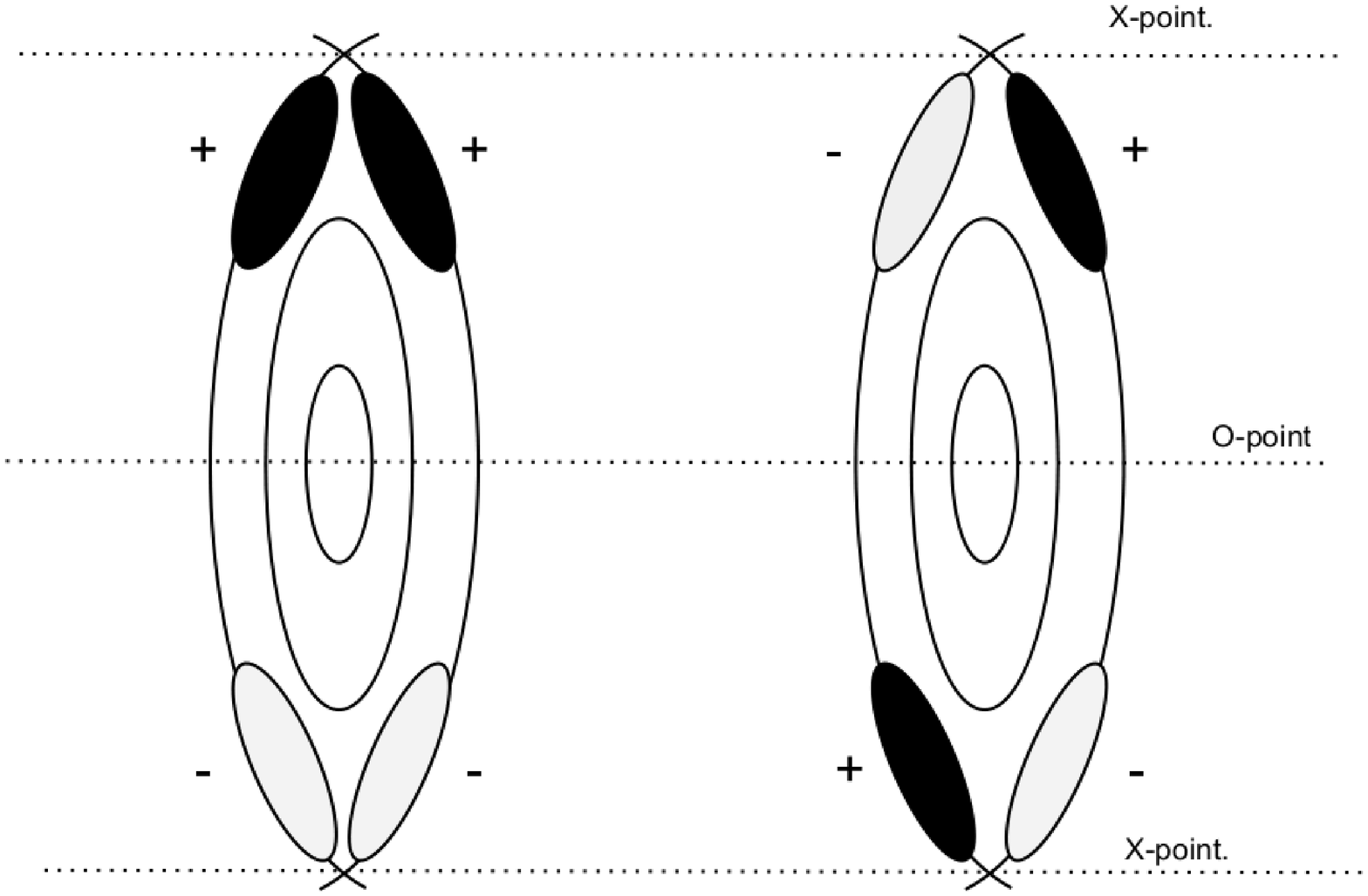}
\caption{Cartoon representing two configurations of density perturbation around magnetic flux surfaces in a magnetic island.  (left) reflected, symmetric solution between inner and outer island regions, producing an up-down density asymmetry and a degree of compression, (right) zero compression perturbation where the outer solution is simply a copy of the inner solution, producing a radially asymmetric density perturbation.}
\label{cartoon}
\end{figure}

Comparison of the island modified GAM dispersion with the result for toroidal circular flux surfaces, 
shows that the oscillations are significantly modified by the presence of a magnetic island.     The oscillation 
has a longer oscillation period that the standard GAM by a factor of approximately $(k^{I}_\theta \rho_{i})^2$, which, is of the order of $10^{-3}$.  Fig.~\ref{eigenfunction}  plots the density 
eigenfunction in the helical island direction. $k^{I}_\theta$ being the islands' poloidal wavevector, which appears whenever a derivative, $\nabla\Omega$, is performed.  Plotted are the four lowest harmonics, higher harmonics exist but are neglected here.
The inlay shows the function in the poloidal angle, which is sinusoidal in the same way as the standard GAM, but of
significantly smaller amplitude.  The function being dominated by the helical direction.

The compression in this case is supplied by the variation of the magnetic field, B, as we travel around the magnetic island.  A net compression exists
when the Eigenfunction is a symmetric reflection between the outer and inner half of the island solutions, which can also be thought of as a symmetric solution in the radial coordinate (See Fig.~\ref{cartoon} for a simple depiction).  When a radially asymmetric solution exists the net
compressive effect is zero and the solution represents a pure sound wave (e.g. Right hand panel of Fig.~\ref{cartoon}).  

Unlike the case of the normal GAM, when the inverse aspect ratio is set to zero 
we obtain solutions that are pure sound waves and the eigenvalue of the symmetric and anti-symmetric solutions are identical.  With a finite aspect ratio 
these two value diverge as the compressive part becomes larger.

The damping rate is determined by kinetic effects as showed by Hinton and Rosenbluth\cite{HinRosPRL,HinRosPPCF},  however we see from 
the time traces of the potential amplitude seen in Fig.~\ref{tracecomp} that the damping rate is significantly slower than the 
rate for the GAM.  

The collisionless damping rate of the GAM, normalised to the ion transit frequency, has been shown to have  the following form\cite{SugGDamp},

\begin{equation}
\frac{\gamma}{k_{\parallel}v_{th i}} \sim \left(\frac{\omega}{\omega_t}\right)^{4}\exp{\left(-\frac{\omega^{2}}{\omega_t}\right)},
\end{equation}
where $\omega$ is the oscillation frequency and $\omega_{t}$ is the ion transit frequency, estimated
by $\omega_t \sim k_{\parallel}v_{th i} \sim k_{\theta}w\hat{s}v_{th i}/q $ in the presence of an island.  With some algebra it can be shown that the ratio of the damping rate to the transit frequency is, $\frac{\gamma_I}{\omega_I} \sim 0.2$.  Where the subscript I denotes the island modified GAM.

With the same analysis; taking the ion transit time for the normal Geodesic Acoustic Mode as, $k_{\parallel} = 1/(R_{0}q)$, we can calculate the damping rate as $\gamma_G \sim 0.36  {v_{th i}/R_0}$.

Finally, the ratio of the normal GAM damping rate to the island modified GAM damping rate is $\frac{\gamma_{G}}{\gamma_{I}} \sim 15$.   . 
From this rough calculation we see that the damping rate of the island modified mode is significantly smaller than the damping rate of the GAM.  

Both modes are evident in the oscillations seen in the trace, the faster GAM, and the slower island oscillation.  We note 
here that the faster oscillation is persistent in the turbulence simulation as it is being continuously excited by the turbulence, while in the 'linear' simulation 
only an initial perturbation is possible which damps accordingly.

\section{Gyro-kinetic framework}
\label{theory}

The oscillatory vortex mode was observed in gyro-kinetic simulations studying the effect of magnetic islands on drift-wave turbulence.  Since the GAM is a linear mode
it should be possible to excite them without turbulence present.

Here the GAM oscillation is studied using the gyro-kinetic framework, with numerical solutions obtained using a modified version of the gyro-kinetic flux-tube code {\small GKW} of which more details can be found in \cite{PEE09}.
The delta-$f$ approximation is used, in which the distribution function is split into a background $F$ and a perturbed 
distribution $f$.
The final equation for the perturbed distribution function $f$, for each species can be written in the form 
\be 
{\partial g \over \partial t} + (v_\parallel {\bf b} + {\bf v}_D) \cdot \nabla f +  {\bf v}_\chi\cdot \nabla g  
-{\mu B \over m}{{\bf B}\cdot \nabla B \over B^2}{\partial f \over \partial v_\parallel} = S, 
\label{gyrovlas}
\ee
where $S$ is the source term which is determined by the background distribution function, $\mu$ is the magnetic moment, $v_{||}$ is the velocity along the magnetic field, $B$ is the magnetic field strength, m and Z are the particle mass and charge number respectively. Here,  $g = f + (Ze/T)v_{\parallel}\langle A_{\parallel} \rangle F_{M}$ is used to absorb the time derivative of the parallel vector potential $\partial A_{\parallel}/\partial t$ which enters the equations through Amp\`{e}res law.  
The background is assumed to be a Maxwellian ($F_M$), with particle density ($n$) and temperature ($T$)
\be 
\label{maxwell}
F = F_M = {n \over \pi^{3/2} v_{\rm th}^3 } \exp \biggl [ - {v_\parallel^2 + 2 \mu B / m \over v_{\rm th}^2}  \biggr ] , 
\ee 
which determines the source term, neglecting temperature and density gradients becomes:
\bee
S = -{Ze \over T} [ v_\parallel {\bf b} + {\bf v}_D ] \cdot \nabla \langle \phi \rangle  F_M . \label{source} 
\eee
\noindent
The thermal velocity $v_{\rm th}\equiv \sqrt{ 2 T / m}$, and  the major radius ($R$) are use to normalise the length and time scales.
Using standard gyro-kinetic ordering, the length scale of perturbations along the field line ($R \nabla_\parallel  \approx 1$)  are significantly longer than those perpendicular to the field ($R \nabla_\perp \approx 1/ \rho_*$). Here, $\rho_* = \rho_i / R$ is the normalised ion Larmor radius (where $\rho_i = m_i v_{th} / e B$ and $v_{th} = \sqrt{2 T_i / m_i}$).

The velocities in Eq.~(\ref{gyrovlas}) are from left to right: the parallel motion along the 
unperturbed field ($v_\parallel {\bf b}$), the drift motion due to the inhomogeneous field 
(${\bf v}_D$), and the motion due to the perturbed electromagnetic field (${\bf v}_\chi$). 
The drift due to the inhomogeneous magnetic field can be written in the form\cite{PEE09},
\be
{\bf v}_D = 
{1\over Ze} \biggl [ {m v_\parallel^2\over B} + \mu \biggr ] {{\bf B} \times \nabla B \over B^2},
\ee
whereas the motion due to the perturbed electromagnetic field 
\be 
{\bf v}_\chi = {{\bf b} \times \nabla \chi \over B},
\ee 
\noindent
 is the combination of the $E \times B$ velocity (${\bf v}_E = {\bf b} \times \nabla \langle \phi \rangle / B$) and the parallel motion along 
the perturbed field line (${\bf v}_{\delta B} = - {{\bf b} \times \nabla v_\parallel 
\langle A_\parallel \rangle /B}$).
These two effects are combined into one velocity through the definition of a new field  $\chi = \langle \phi \rangle - v_\parallel \langle A_\parallel \rangle$.  
Here, the angled brackets denote gyro-averaged quantities.

The electrostatic potential is calculated from the gyro-kinetic Poisson equation which in Fourier space is
\bee 
\label{Poisson}
 \sum_{sp}  Z_{sp} n_{Rsp} \biggl [ 2 \pi B \int {\rm d} v_{\parallel} {\rm d} \mu J_0(k_\perp\rho_{sp}) \hat g_{sp} + \nonumber\\
{Z_{sp} \over T_{Rsp}} [ \Gamma(b_{sp}) -1]\hat \phi \biggr ] = 0 , 
\eee
where  
$ b = {1 \over 2} m_R T_R (k_\perp \rho_* R_{\rm ref} / Z B^2)^2  =  {1 \over 2} {k_\perp^2 m^2 v_{\rm th}^2 \over Z^2e^2 B^2} $, $k_\perp$ being the perpendicular wave-number and $J_0$ are zeroth order Bessel functions of the first kind.

\begin{figure}
\centering
\includegraphics[width=9.0cm,clip]{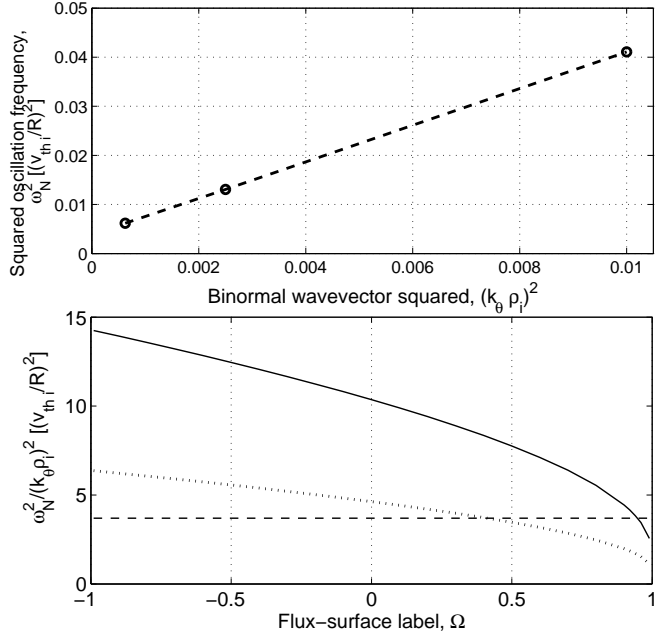}
\caption{(top) The square of the oscillation frequency for individual code runs where the value of $k_{\theta}\rho_{i}$, 
and hence the toroidal size of the island, is varied, plotted against the square of the wave-vector.  The dashed line represents a linear fit.  
This shows an exactly linear dependence.  (Bottom) The calculated ratio of the Eigenvalue and $(k_{\theta}\rho_i)^{2}$  as a function of the perturbed 
flux label $\Omega$ from the O-point $\Omega=-1$ to the separatrix, $\Omega=1$.  The agreement between the calculated gradient from the (top) figure, 
$\omega_{N}^{2}/(k_{\theta}\rho_{i})^{2} = 3.7$ (represented by the dashed line, here $\omega_{N} = \omega_I v_{th i}/R$) compares well with the value at approximately the island separatrix,
 $\Omega\sim0.9$.  The code was run with 50 points in both the $\xi$ and $\theta$ directions.}
\label{disprelation}
\end{figure}

GKW uses straight field line Hamada\cite{HAM58} coordinates ($s,\zeta,\psi$) where $s$ is the coordinate along the magnetic field and $\zeta$ is the generalised toroidal angle. For circular concentric surfaces, the transformation of poloidal and toroidal angle to these coordinates is given by  \cite{PEE09} ($s,\zeta) = (\theta / 2 \pi , [q \theta - \phi]/ 2 \pi)$.
Assuming the winding of the magnetic field is resonant ($q = m/n$) in the centre of the computational domain ($\psi_{0} = r_{0}/R_{0} = \epsilon$, where $r$ is the radius of the magnetic surface, and $R_{0}$ is the distance of the centre of the surface to the axis of symmetry) and  expanding $q$ up to first order in $\Delta\psi$ ($\Delta\psi$ being the radial distance from the resonant surface and here m is the poloidal mode number), $q = m / n + \Delta\psi (\partial q /\partial\psi)$ then yields
\begin{equation}
A_\parallel = \tilde A_{\parallel} \exp [ 2 \pi {\rm i} n (\zeta - s{\partial q / \partial \psi}\Delta\psi)].
\end{equation}
The wave vector of the island is 
\begin{math}
k_\zeta^{I} \rho_i = 2 \pi n \rho_*.
\end{math}
GKW uses a Fourier representation in the plane perpendicular to the magnetic field. The periodicity constraint on the torus shaped magnetic surface then dictates a relation between the radial and toroidal modes.

The half width of the island is defined by
\be
w = 2\sqrt{ {q \tilde{\psi} / \hat s RB }},  
\ee
(where $\hat{s} = (1/q)\partial q/\partial \psi$ is the magnetic shear) and the perturbed magnetic flux, $\tilde{\Psi}$, is related to the perturbation of the parallel vector potential by the relation,
\be
\tilde{\psi}= -RA_{\parallel}
\ee

Full details of the numerical implementation of the magnetic island is omitted here,  the interested 
reader can find them in these papers \cite{Hor10,POL10}.

\section{Results and comparison}
\label{compare}

Presented here are the results from simulations, which keep the kinetic electron effects with the true mass ratio of a Deuterium plasma.   While we are studying the linear response to a perturbation, due to the set-up of GKW, the code must be run non-linearly for the plasma to feel the effects of the modified field lines due to the magnetic island.  
This is because, due to numerical reasons, the parallel vector potential of the island is introduced as a perturbation.

The parameters used for these simulations are similar (but not equivalent) to those of the cyclone base case \cite{DIM00}.  However, to simplify the physics we have set the temperature and density gradients in the background distribution ($R/L_T= R/L_N=0$)to zero, otherwise : 
\begin{itemize}
\item Inverse aspect ratio $\epsilon = 0.19$
\item Electron to ion temperature ratio $T_e / T_i = 1$
\item Safety factor $q = 1.5$ and magnetic shear $\hat s = 0.16$.
\item 2 toroidal modes, 167 radial modes.  Results are presented with, $k_\zeta^{I} \rho_i = 0.025$, $k_\zeta^{I} \rho_i = 0.05$, $k_\zeta^{I} \rho_i = 0.1$
\end{itemize}

\begin{figure}
\centering
\includegraphics[width=8.5cm,clip]{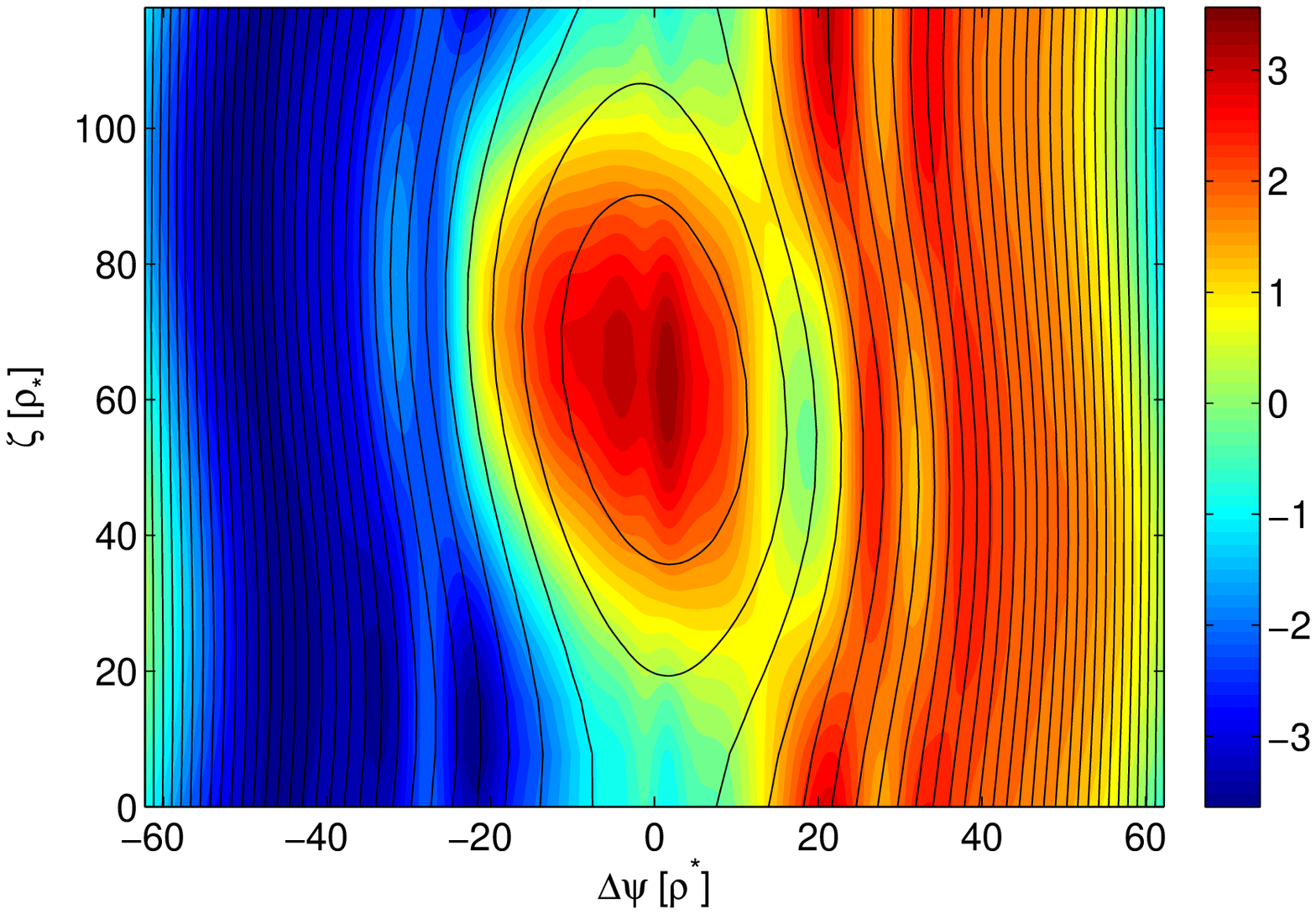}
\includegraphics[width=8.5cm,clip]{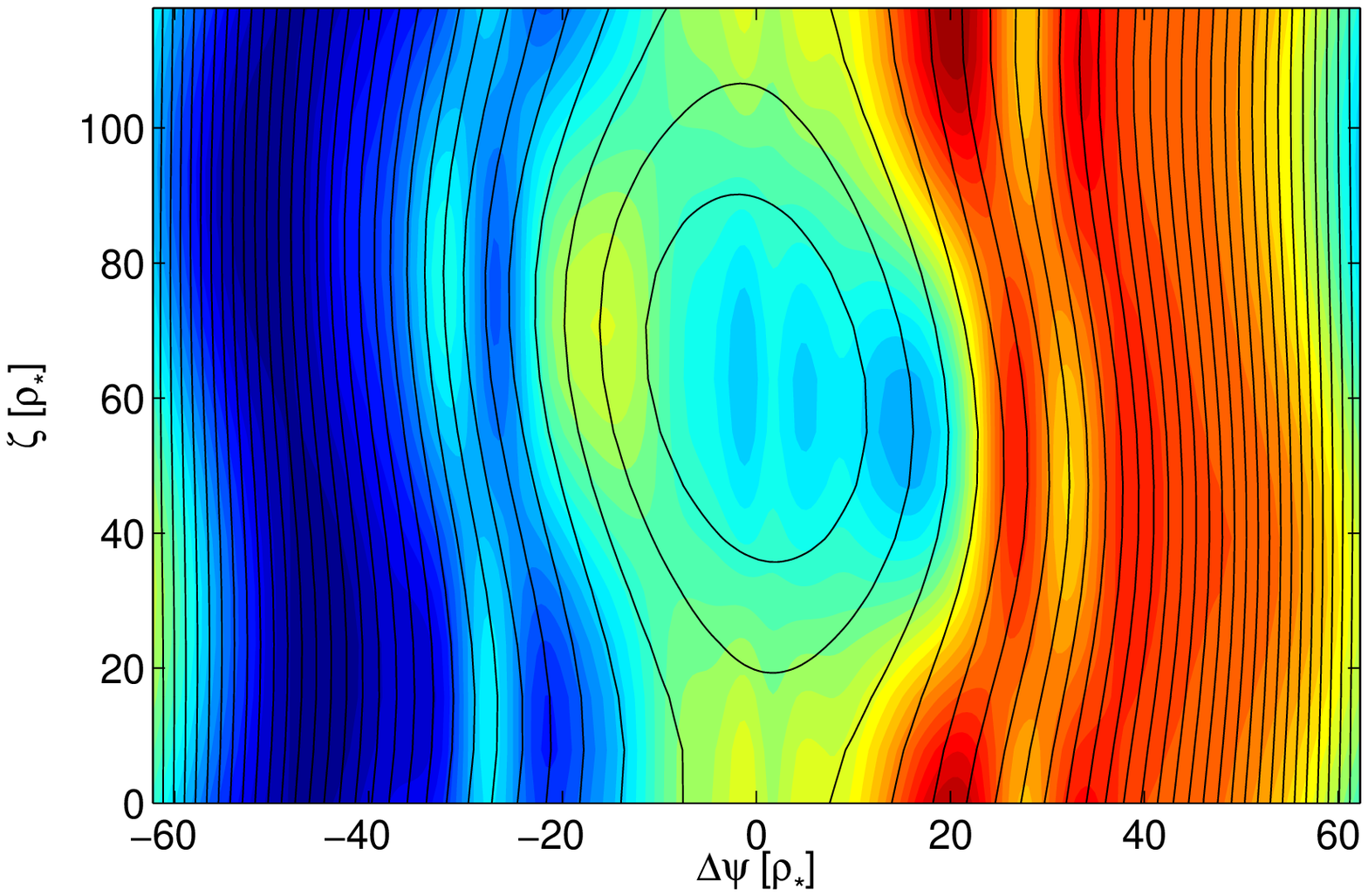}
\caption{(Color online) Normalized electrostatic potential ($\phi_{N}=e\phi/T\rho_*$, $W/\rho_{i}=24$) in the plane perpendicular to the magnetic field (outboard mid-plane).  Black lines represent the perturbed flux surfaces calculated from the total parallel vector potential.  The top panel shows a vortex with a positive sign, while the lower panel shows that the vortex has flipped sign at a point later in the simulation.  Data from Gyro-kinetic simulation without turbulence and initialised
with a $\sin\xi$ density perturbation which produces a electrostatic vortex structure similar to those seen in turbulent Gyro-kinetic simulations (See Fig.~\ref{islpotturb}.)}
\label{islpot}
\end{figure}

For an idea of the typical size of an island, consider a $m = 3$, $n = 2$ island which is resonant at $q = 1.5$.  This choice effectively determines $\rho_* = 4 \cdot 10^{-3}$ in the  $k_\zeta \rho_i = 0.05$  case and $\rho_* = 2 \cdot 10^{-3}$ in the $k_\zeta \rho_i = 0.025$  case, values that correspond to a medium-size tokamak such as ASDEX Upgrade\cite{ASDU}. 

Run in this way, the code encapsulates all the necessary physics of the Geodesic acoustic mode in the MHD limit.  
A density perturbation is initialised which is resonant with the magnetic island, $\rho=\rho_{0}\sin\xi$ and 
then allowed to freely evolve.    An oscillatory vortex mode is excited, as seen in Fig~\ref{islpot}.  The initial perturbation, however, is not an exact eigenfunction of the system and thus we get some extra radial oscillations (See Fig.~\ref{islpot}) that damp away on a long time scale.

Fig.~\ref{tracecomp} shows the time trace of the amplitude of the first non-zero $k_{x}$ electrostatic potential 
mode which has the same poloidal mode number as the magnetic island.  The trace compares the potential from a gyrokinetic simulation (dashed line) without turbulence and one with (black line).

From these traces we see that in both cases there is indeed two frequencies present.  The first, faster component is the 
standard GAM oscillation produced by the Geodesic curvature of the circular flux surfaces which are also damped 
by kinetic effects\cite{HinRosPPCF}.   The second, significantly longer period and higher amplitude 
oscillation, is the Geodesic mode around the closed flux surfaces within the magnetic island separatrix.  
This is slowly damped compared with the usual GAM case.  There is indeed some disparity in the frequencies between 
the linear and nonlinear turbulent simulation traces, however the physics is significantly different between 
the two, with turbulence and the presence of equilibrium temperature and density gradients in the non-linear 
case, which could have a significant impact on the frequency of the vortex mode.

Plotted in Fig.~\ref{disprelation} is the squared oscillation frequency  against the normalised squared toroidal wave-vector associated with the magnetic island, $k_{\theta}^{I} = m/r$.  We see that there is an exactly linear relation between these parameters.  In the
lower panel is plotted the squared frequency as a function of the flux surface label within the magnetic island as calculated from the eigenvalue solver with the same parameters as used
in the gyro-kinetic simulations.
Agreement between the simulations and the eigenvalue analysis is very good, with the frequency values matching near to the separatrix  ($\Omega=0.95$).  An analysis with a kinetic or higher order theory which takes into account variation of the mode across the flux surfaces would give a more accurate calculation of the oscillation frequency, but it beyond the scope of this paper.

It is observed that the mode from our eigenvalue analysis, with the closest matching frequency to that seen in gyro-kinetic turbulence simulations is the one which corresponds to an up-down density asymmetry within the magnetic island, with an eigenfunction corresponding to the full
line in Fig.~\ref{eigenfunction} and its symmetric reflection.  This is depicted in cartoon form in the left panel of Fig.~\ref{cartoon}.  Solutions of this form have a GAM component, and therefore introduce a compression.  The compression produces an electric field that is perpendicular to the flux surfaces, and is therefore essential in 
producing the vortex structures as seen in Fig.~\ref{islpot}.  

Indeed, solutions exist which have good agreement with between their eigenvalue and the  frequency observed in gyrokinetic simulations, however, these have no
GAM component and are therefore purely sound waves.  Solutions of this form are unable to produce the vortex structures observed as plotted in Fig.~\ref{islpot} (and its electrostatic potential time trace in Fig.~\ref{tracecomp}).  One example of this is the eigenfunction represented by the dot-dashed line in Fig.~\ref{eigenfunction} whose eigenvalue is plotted (dotted line) in the lower panel of  Fig.~\ref{disprelation}.  

In turbulence simulations it was observed that spreading occurred and turbulent structures entered the island from the upper x-point and spread down into the island, giving an up down asymmetry (See Fig.5 in Ref.~\cite{Hor10}).  
It is this mechanism which is a candidate to excite the oscillatory mode seen that has a value close to the calculated frequency near the separatrix and not a higher frequency as would be expected further toward the O-point.

\section{Conclusions}

Motivated by the observation of oscillatory potential vortex structures seen in simulations of turbulence 
around magnetic islands, we have performed an analysis of the Geodesic acoustic mode around a magnetic island.  
It is seen that long time-scale oscillatory solutions are generated with the same properties as the Geodesic acoustic mode, namely 
plasma compressibility producing an electric field perpendicular to the perturbed flux surfaces, which produce meso-scale vortex structures.

Also performed were Gyro-kinetic simulations where the turbulence was neglected and the density initialised to be present with the magnetic island, which 
generates an oscillatory potential structure, which is the same as that seen in turbulence simulations.  The scaling of the frequency of this oscillation agrees with our 
eigenvalue analysis, leading us to conclude that the oscillatory structures are indeed 
the Geodesic Acoustic Mode around the closed flux surfaces within the magnetic island.

These oscillatory vortex structure can induce $E\times B$ flow around large magnetic islands which 
have a profound effect heat transport in the vicinity of the magnetic island and can also have a 
significant regulatory effect on the turbulence in this region.

\begin{acknowledgments}
This work used resources on the HECToR supercomputer that were
provided by the Engineering and Physical Sciences Research Council
[grant number EP/H002081/1].
\end{acknowledgments}

\section{Appendix A - Treatment of ExB term}
\label{EcrossB}

Firstly we consider the denominator of the compressive term, $\int \frac{\left|\nabla\Omega\right|^2}{B^2}JdS$.  If we utilise:
\begin{equation}
\nabla\Omega = \frac{4(r-r_0)}{w^2}\hat{\mathbf{r}} + \frac{m}{r}\sin{\xi}\hat{\mathbf{\theta}} - \frac{n}{R}\sin{\xi}\hat{\mathbf{\zeta}}
\end{equation}

We utilise the magnetic field approximation $B^2 = B_0^2 /(1 + \epsilon\cos\theta)^2$ and Taylor expand, to give:
\begin{eqnarray}
&\int& \frac{\left|\nabla\Omega\right|^2}{B^2}JdS = \\ 
&\int& JdS \frac{d\xi}{\sqrt{\Omega + \cos\xi}}  \frac{1}{B_0^2}\left( \frac{16(\Omega + \cos\xi)}{w^2}\right)(1+2\epsilon\cos\theta) \nonumber\\ 
&+& \int JdS \frac{d\xi}{\sqrt{\Omega + \cos\xi}}  \frac{1}{B_0^2}\left( \frac{m^2}{r^2} + \frac{n^2}{R^2} \right)(1+2\epsilon\cos\theta)\sin^2{\xi} \nonumber
\end{eqnarray}
We first perform the $\theta$ integral, which removes the terms of order $\epsilon$, we obtain,
\begin{eqnarray}
\frac{2\pi}{B_0^2 w^2}\Bigg( 16 \int \frac{(\Omega + \cos\xi)}{\sqrt{\Omega + \cos\xi}} d\xi + w^2\Bigg( \frac{m^2}{r^2} + \nonumber\\
\frac{n^2}{R^2} \Bigg)\int \frac{\sin^2\xi}{\sqrt{\Omega + \cos\xi}} d\xi\Bigg).
\end{eqnarray}

We assume a small aspect ratio $(r<R)$ and also small island width in relation to the minor radius$(w<r)$ and as such the integral reduces to:
\begin{equation}
\frac{16\pi}{B_0^2 w^2} \int \frac{(\Omega + \cos\xi)}{\sqrt{\Omega + \cos\xi}} d\xi
\label{part3}
\end{equation}

Finally we treat the term $\int \tilde{\rho} \frac{\mathbf{B}\times\nabla \Omega.\nabla B^{2}}{B^{4}} JdS$, utilising $\mathbf{B} = B_{t}\hat{\mathbf{\zeta}} + B_{\theta}\hat{\mathbf{\theta}} + \frac{\tilde{\psi}m}{Rr}\sin\theta \hat{\mathbf{r}}$ and also:
\begin{eqnarray}
\nabla B^2  &=& 2B\nabla B  \nonumber\\
&=& -\frac{2B_0^2}{R(1+\epsilon\cos\theta)^{3}}\left(\cos\theta \mathbf{\hat{r}} - \sin\theta\mathbf{\hat{\theta}}\right) 
\end{eqnarray}
\noindent
Taking the cross product with the magnetic field vector:
\begin{eqnarray}
\mathbf{B}\times\nabla B^{2}&=&B_t\hat{\zeta}\times\nabla B^2 + B_{\theta}\hat{\theta}\times\nabla B^2 + \nabla\zeta\times\nabla\psi\times\nabla B^2 \nonumber\\
&=& -\frac{2B_{0}^{3}}{R(1+\epsilon\cos\theta)^{3}}\Bigl( B_{t}\cos\theta \hat{\mathbf{\theta}} - B_{\theta}\cos\theta \hat{\mathbf{\zeta}}\nonumber\\ 
&+& B_{t}\sin\theta \hat{\mathbf{r}} - \frac{\tilde{\psi}m}{Rr}\sin\theta\sin\xi\hat{\mathbf{\zeta}} \Bigr)
\label{crossprod}
\end{eqnarray}
\noindent
Taking the inner product with $\nabla\Omega$, this gives four terms:
\begin{eqnarray}
\mathbf{B}\times\nabla B^{2}\cdot\nabla\Omega&=& \frac{2B_{0}^{3}}{R(1+\epsilon\cos\theta)^{3}}\Bigl(-\frac{4(r-r_{0})}{w^{2}}B_{t}\sin\theta \nonumber\\ 
&-&\frac{B_{t}m}{r}\cos\theta\sin\xi -\frac{B_{\theta}n}{R}\cos\theta\sin\xi \nonumber\\
&-&\frac{mn\tilde{\psi}}{rR}\sin^{2}\xi\sin\theta  \Bigr)
\label{compre}
\end{eqnarray}
We make the approximations, neglecting the effect of the island on the field strength:
\begin{eqnarray}
B_t &=& B_0/(1+\epsilon\cos\theta)\nonumber\\
B_\theta &=& \frac{\epsilon B_0}{q}/(1 + \epsilon\cos\theta) 
\end{eqnarray}
we finally obtain:
\begin{eqnarray}
\frac{\mathbf{B}\times\nabla B^{2}\cdot\nabla\Omega}{B^{4}}&=& \frac{-2}{B_{0}R}\Bigl(\frac{4(\Omega+\cos\xi)}{w\sqrt{2}}\sin\theta \nonumber\\ 
&+&\frac{m}{r}\cos\theta\sin\xi +\frac{\epsilon n}{qR}\cos\theta\sin\xi \nonumber\\
&-&\frac{mn\tilde{\psi}}{rR}\sin^{2}\xi\sin\theta(1+\epsilon\cos\theta)  \Bigr)
\label{comprefin}
\end{eqnarray}
The last term in Eq.~(\ref{comprefin}) can be neglected as it is comparably small with respect to the other terms.


\begin{thebibliography}{10}

\bibitem{FUR63} H.~P.~Furth, J.~Killeen, M.~N.~Rosenbluth, Phys. Fluids {\bf 6} 459 (1963)

\bibitem{RUTH73} P.~H.~Rutherford, Phys. Fluids {\bf 16} 1903 (1973)

\bibitem{WAE09} F.~L.~Waelbroeck, Nucl. Fusion {\bf 49} 104025 (2009)


\bibitem{DiaZF} P.~H.~Diamond, S.~I.~Itoh, K.~Itoh, Plasma Phys. Contol. Fusion {\bf 47} R35-R161 (2005)

\bibitem{HasZF} A.~Hasegawa, C.~G.~Maclennan, Y.~Kodama, Phys. Fluids {\bf 22} 2122 (1979)

\bibitem{WalGF} R.~E.~Waltz, C.~Holland, Phys. Plasmas {\bf 15} 122503 (1994)

\bibitem{LinZF} Z.~Lin, T.~S.~Hahm, W.~W.~Lee, W.~M.~Tang, R.~B.~White, Science {\bf 18} 1835 (1998)
 

\bibitem{WangVort} Z.~X.~Wang, J.~Q.~Li, J.~Q.~Dong, and Y.~Kishimoto, Phys. Plasmas {\bf 18} 012110 (2011)

\bibitem{WangFlow} Z.~X.~Wang, X.~Wang, J.~Q.~Dong, and Y.~Kishimoto, and J.~Q.~Li, Phys. Plasmas {\bf 15} 082109 (2008)

\bibitem{Mura09} M.~Muraglia, O.~Agullo, M.~Yagi, S.~Benkadda, P.~Beyer, X.~Garbet S.~-I.~Itoh, K.~Itoh and A.~Sen, Nucl. Fusion, {\bf 49} 055016


\bibitem{HorEPL} W.~A.~Hornsby, A.~G.~Peeters, E.~Poli, M.~Siccinio, A.~P.~Snodin, F.~J.~Casson, Y.~Camenen, G.~Szepesi, Euro. Phys. Lett. {\bf 91} 45001 (2010)

\bibitem{POL09} E.~Poli, A.~Bottino and A.~G.~Peeters, Nucl. Fusion {\bf 49}, 075010 (2009)

\bibitem{POL10} E. Poli, A.~Bottino, W.~A.~Hornsby, A.~G.~Peeters, T.~Ribeiro, B.~D.~Scott, M.~Siccinio, Plasma Phys. Control. Fusion {\bf 52} 124021 (2010)


\bibitem{Hor10} W.~A.~Hornsby, A.~G.~Peeters,  A.~P.~Snodin, F.~J.~Casson, Y.~Camenen, G.~Szepesi, M.~Siccinio, E.~Poli, Phys. Plasmas {\bf 17} 092301 (2010)

\bibitem{HorVar} W.~A.~Hornsby, M.~Siccinio, A.~G.~Peeters, E.~Poli,  A.~P.~Snodin, F.~J.~Casson, Y.~Camenen, G.~Szepesi, Plasma. Phys. Control. Fusion {\bf 53 } 054008 (2011)

\bibitem{CAR86} R. Carrera, R.D. Hazeltine, M. Kotschenreuther, Phys. Fluids {\bf 29} 899 (1986)

\bibitem{WIL96} H.R. Wilson, J.W. Connor, R.J. Hastie, and C.C. Hegna, Phys. Plasmas {\bf 3} 248 (1996)



\bibitem{Win68} N. Windsor, J.L. Johnson, J.M. Dawson, Phys. Fluids {\bf 11} 2248 (1968)

\bibitem{Hag09} R.~Hager and K.~Hallatschek, Phys. Plasmas {\bf 16} 072503 (2009)

\bibitem{Shi05} B.~Shi, J.~Li and J.~Dong, Chin. Phys. Lett. {\bf 22} 1179 (2005)

\bibitem{QiuZon} Z.~Qiu, F.~Zonca and L.~Chen, Plasma Sci. and Tech. {\bf 13} 3 (2011)


\bibitem{ConwayGAM05} G.D. Conway, B. Scott, J. Schirmer, M. Reich, A. Kendl and the ASDEX Upgrade Team, Plasma Phys. Control. Fusion {\bf 47} 1165 (2005)

\bibitem{HamadaGAM05} Y. Hamada, A. Nishizawa, T. Ido, T. Watari, M. Kojima, Y. Kawasumi, K. Narihara, K. Toi and JIPPT-IIU Group, Plasma Phys. Control. Fusion {\bf 45} 81 (2005)

\bibitem{MelnikovGAM06} A. V. Melnikov, Plasma Phys. Control. Fusion {\bf  48} S87 (2006)

\bibitem{ZhaoGAM06} K.~J.~Zhao, T.~Lan, J.~Q.~Dong, L.~W.~Yan, W.~Y.~Hong, C.~X.~Yu, A.~D.~Liu, J.~Qian, J.~Cheng, D.~L.~Yu, Q.~W.~Yang, X.~T.~Ding, Y.~Liu, and C.~H.~Pan, Phys. Rev. Lett. {\bf 96} 255004 (2006)

\bibitem{FujisawaGAM07} A.~Fujisawa, T.~Ido, A.~Shimizu, S.~Okamura, K.~Matsuoka, H.~Iguchi, Y.~Hamada, H.~Nakano, S.~Ohshima, K.~Itoh, K.~Hoshino, K.~Shinohara, Y.~Miura, Y.~Nagashima, S.-I.~Itoh, M.~Shats, H.~Xia, J.~Q.~Dong, L.~W.~Yan, K.~J.~Zhao, G.~D.~Conway, U.~Stroth, A.~V.~Melnikov, L.~G.~Eliseev, S.~E.~Lysenko, S.~V.~Perfilov, C.~Hidalgo, G.~R.~Tynan, C.~Holland, P.~H.~Diamond, G.~R.~McKee, R.~J.~Fonck, D.~K.~Gupta and P.~M.~Schoch, Nucl. Fusion {\bf 47} S718 (2007) 

\bibitem{WatariGAM06} T. Watari, Y. Hamada, T. Notake, N. Tekeuchi and K. Itoh, Phys. Plasmas {\bf 13} 062504 (2006)

\bibitem{WatariGAM05} T. Watari, Y. Hamada, A. Fujisawa, K. Toi and K. Itoh, Phys. Plasmas {\bf 12} 062304 (2005)

\bibitem{SmolGAM} A.~I.~Smolyakov, X.~Garbet and M.~Ottaviani, Phys. Rev. Lett. {\bf 99} 055002 (2007)

\bibitem{biancalani} A.~Biancalani and L.~Chen and F.~Pegoraro and F.~Zonca, Phys. Rev. Lett. {\bf 105} 095002 (2010)

\bibitem{FIT95} R. Fitzpatrick, Phys. Plasmas {\bf 2}, 825 (1995)

\bibitem{WIL09} H.R. Wilson, and J.W. Connor, Plasma Phys. Contr. Fusion {\bf 51}, 115007 (2009)

\bibitem{sicc09} M.~Siccinio and E.~Poli, Plasma Phys. Control. Fusion {\bf 51} 075005 (2009)

\bibitem{jame} M.~James and H.~R.~Wilson, Plasma Phys. Contol. Fusion {\bf 48} 1647-1659 (2006)

\bibitem{smol95} A.~I.~Smolyakov, A.~Hirose, E.~Lazzaro, G.~B.~Re and J.~D.~Callen, Phys. Plasmas {\bf 2}, 1581 (1995)


\bibitem{sug06} H.~Sugama and T.~-H.~Watanabe, J. Plasma Physics {\bf 72} 6 (2006)


\bibitem{HinRosPRL} M.~N.~Rosenbluth  and F.~L.~Hinton, Phys. Rev. Lett. {\bf 80} 724-727 (1998)

\bibitem{HinRosPPCF} F.~L.~Hinton and M.~N.~Rosenbluth, Plasma Phys. Control. Fusion {\bf 41} A653 (1999)

\bibitem{SugGDamp} H.~Sugama and T.~H.~Watanabe, J. Plasma Physics {\bf 72} 825 (2006)

\bibitem{PEE09} A.G. Peeters, Y. Camenen, F.J. Casson, W.A. Hornsby, A.P. Snodin, D. Strintzi, and G. Szepesi, Comp. Phys. Comm. {\bf 180}, 2649 (2009) 

\bibitem{HAM58} S. Hamada, Kakuyugo Kenkyu {\bf 1}, 542 (1958)


\bibitem{DIM00}  A.M. Dimits, G. Bateman, M.A. Beer, B.I. Cohen, W. Dorland, G W. Hammett, C. Kim, J.E. Kinsey, M. Kotschenreuther, A.H. Kritz, L.L. Lao, J. Mandrekas, W.M. Nevins, S.E. Parker, A.J. Redd, Phys. Plasmas {\bf 7}, 3 (2000)

\bibitem{ASDU} ASDEX Upgrade Team, Nucl. Fusion {\bf 39} 1321 (1999)

\end{thebibliography}
\end{document}